\documentclass[%
 reprint,
superscriptaddress,
 amsmath,amssymb,
 aps, braket,
]{revtex4-1}

\usepackage{graphicx}% Include figure files
\usepackage{dcolumn}% Align table columns on decimal point
\usepackage{bm}% bold math
\usepackage{physics}
\usepackage{multirow}
\usepackage[hidelinks]{hyperref}
\usepackage{amsthm}
\usepackage{apptools}
\usepackage{chngcntr}
\usepackage{mathtools}
\usepackage{subcaption}
\usepackage{color}
\AtAppendix{\counterwithin{theorem}{section}}
\makeatletter
\def\blfootnote{\xdef\@thefnmark{}\@footnotetext}
\makeatother

\begin{document}

\preprint{APS/123-QED}

\title{Fault-Tolerant Weighted Union-Find Decoding on the Toric Code}% Force line breaks with \\

\author{Shilin Huang}
\email{shilin.huang@duke.edu}
\affiliation{Departments of Electrical and Computer Engineering, Chemistry, and Physics, Duke University, Durham, NC, 27708, USA}
\author{Michael Newman}
\email{michael.newman@duke.edu}
\affiliation{Departments of Electrical and Computer Engineering, Chemistry, and Physics, Duke University, Durham, NC, 27708, USA}
\author{Kenneth R. Brown}
\email{kenneth.r.brown@duke.edu}
\affiliation{Departments of Electrical and Computer Engineering, Chemistry, and Physics, Duke University, Durham, NC, 27708, USA}

\date{\today}% It is always \today, today,
\begin{abstract}
Quantum error correction requires decoders that are both accurate and efficient.  To this end, union-find decoding has emerged as a promising candidate for error correction on the surface code.  In this work, we benchmark a weighted variant of the union-find decoder on the toric code under circuit-level depolarizing noise.  This variant preserves the almost-linear time complexity of the original while significantly increasing the performance in the fault-tolerance setting.  In this noise model, weighting the union-find decoder increases the threshold from $0.38\%$ to $0.62\%$, compared to an increase from $0.65\%$ to $0.72\%$ when weighting a matching decoder.  Further assuming quantum non-demolition measurements, weighted union-find decoding achieves a threshold of $0.76\%$ compared to the $0.90\%$ threshold when matching.  We additionally provide comparisons of timing as well as low error rate behavior.
\end{abstract}

\maketitle

\section{Introduction}

In order to realize scalable quantum computing, quantum information must be protected in quantum error correcting codes.  Information about the errors occurring are rapidly extracted through measurements, and this information is processed through a decoder in order to determine which errors have occurred.  These decoders must be accurate in providing good estimates for the error, but they should also be highly efficient in order to keep up with the quantum computation as it progresses.

One of the leading candidates for quantum error correction is the surface code \cite{Dennis:2002, kitaev2003fault, fowler2012surface}, owing to its $2$D nearest-neighbor implementation \cite{raussendorf2007fault}, robust memory \cite{fowler2009high}, optimized logical gates \cite{brown2017poking, gidney2019efficient, litinski2019game}, and wealth of decoding schemes \cite{duclos2010fast,fowler2012topological,fowler2012towards,fowler2012timing,suchara2015leakage,wootton2012high,duclos2013fault,hutter2014efficient,bravyi2014efficient,wootton2015simple, fowler2015minimum, watson2015fast, varsamopoulos2017decoding, herold2017cellular, kubica2019cellular, torlai2017neural,tuckett2019tailoring,tuckett2020fault,landahl2011fault, wang2010graphical, wang2011surface, bombin2012universal,chamberland2018deep, li20192d,maskara2019advantages,nickerson2019analysing, delfosse2017almost, huang2019fault}.  Among these schemes, decoding based on minimum-weight perfect matching (MWPM) is particularly promising due to its high performance, adaptability to circuit-level errors, and relative $O(n^3)$ efficiency \cite{edmonds2009paths}.  In particular, there has been significant effort aimed at accelerating and parallelizing MWPM \cite{fowler2012towards, fowler2012timing, fowler2015minimum}.

However, performing decoding at the clock speed of a quantum computer remains a daunting task.  A new type of decoder based on the union-find (UF) primitive has been proposed as an alternative to MWPM \cite{delfosse2017almost}.  This decoder relies on generating an erasure consistent with the syndrome information, and then applying a highly efficient erasure decoder \cite{delfosse2017linear}.  Moreover, the UF decoder remains competitive with the high performance of MWPM in a phenomenological error model \cite{delfosse2017almost, nickerson2018measurement, newman2019generating}.

In this work, we benchmark the UF decoder in the fault-tolerance setting under standard circuit-level depolarizing noise.  We show that by adapting the decoder to weighted graphs, the performance increases substantially.  This variant was first proposed in \cite{huang2019fault}, however, it can be modified to preserve the almost-linear run time of the original UF decoder. 

Weighting the decoder graph is a natural step that yields significant gains in the context of MWPM \cite{wang2011surface}.  In particular, for a properly weighted graph, MWPM decides on the most likely error given a particular syndrome \cite{Dennis:2002, wang2003confinement, wang2011surface}.  While UF decoding does not have a simple interpretation on weighted graphs, it is reasonable to expect that preferencing cluster growth in the direction of the most likely nearby error would be beneficial.  What is remarkable is the degree to which it helps, with a significantly greater relative gain than weighted matching over unweighted matching.

\section{Weighted Union-Find} \label{wuf}

We follow the prescription of the original UF decoder described in \cite{delfosse2017almost}, but with weighted edges on the decoder graph.  The complexity of the original algorithm is dominated by the union-find primitive, which has complexity $O(n\alpha(n))$ \cite{tarjan1975efficiency}, where $\alpha$ is the inverse Ackermann's function and $n$ is the number of syndrome bits.  For all practical sizes, this is essentially linear in $n$ with a small constant.  For fault-tolerant decoding in a distance $d$ toric code, $n = 2d^2$ when averaged over $\propto d$ rounds of syndrome extraction.  This approach straightforwardly generalizes to the open boundaries of the surface code, but we benchmark using periodic boundaries to minimize finite-size effects.

\begin{figure}[htb!]
\begin{subfigure}{\linewidth}
  \centering
  \includegraphics[width=.54\linewidth]{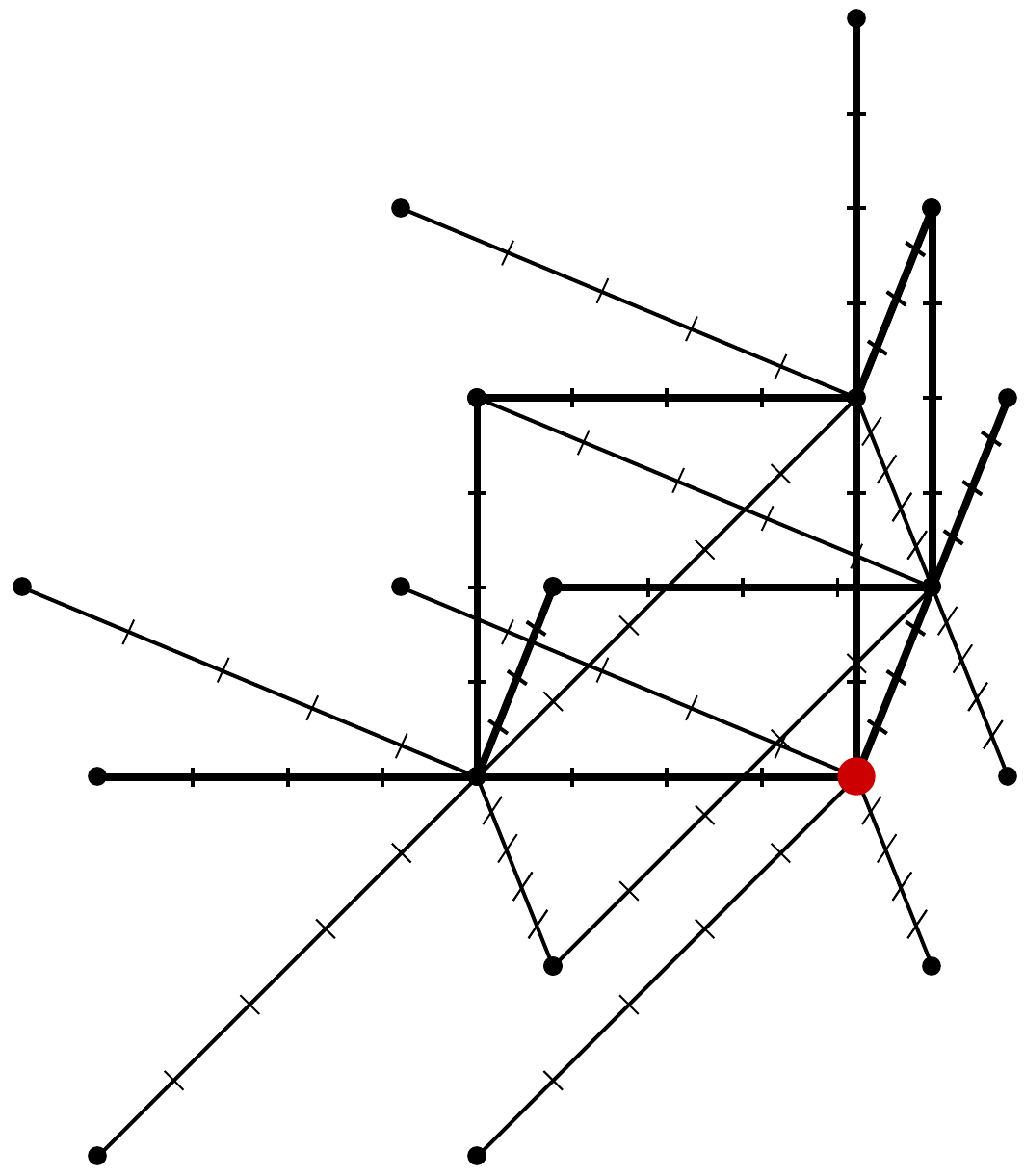}  
  \caption{A single excitation occurs in the corner of the decoder graph.}
  \label{fig:step1}
\end{subfigure}
\begin{subfigure}{\linewidth}
  \centering
  \includegraphics[width=.54\linewidth]{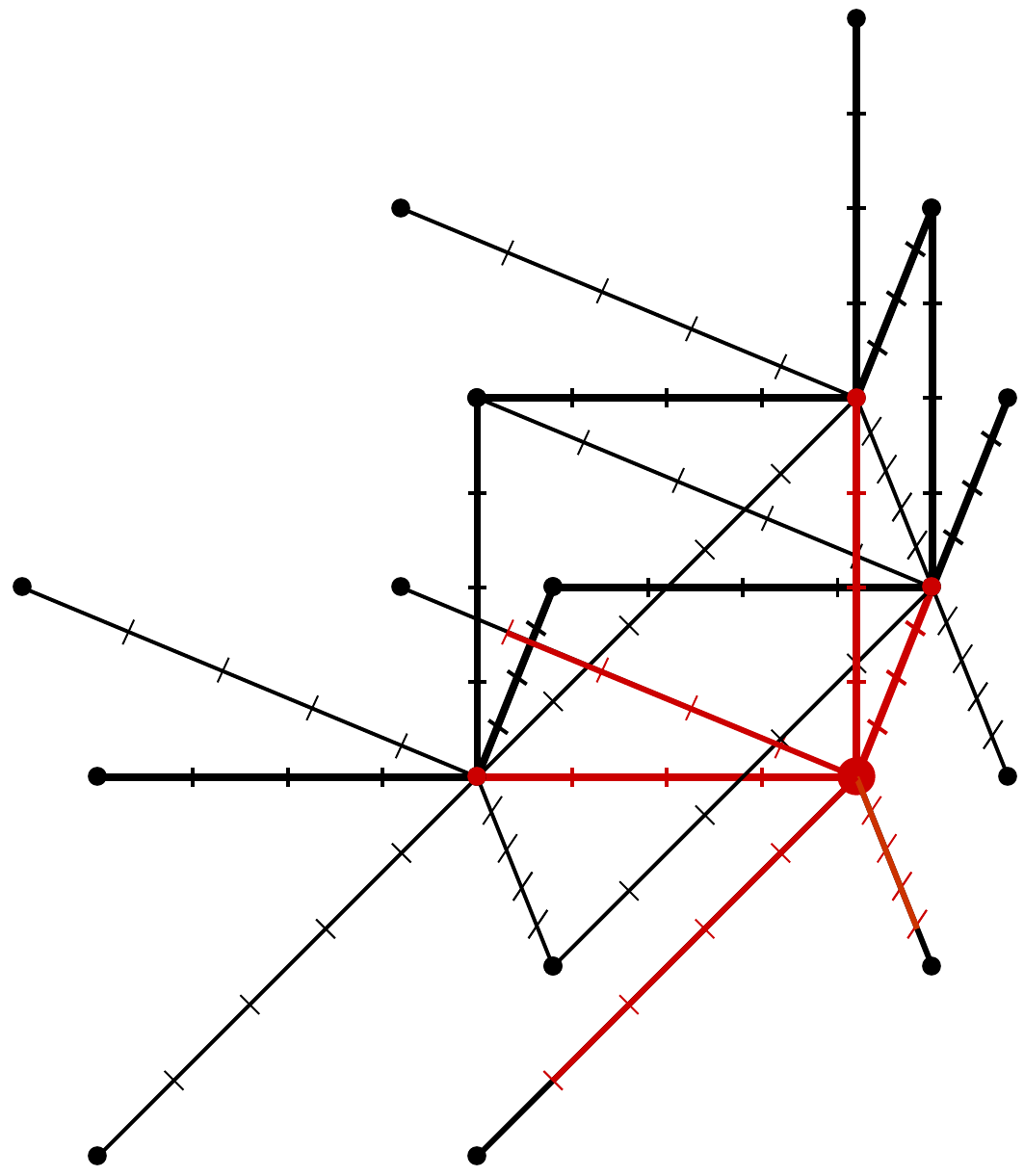}  
  \caption{In the first step, the cluster radius grows by four and it merges with clusters to the north, west, and up directions.}
  \label{fig:step2}
\end{subfigure}
\begin{subfigure}{\linewidth}
  \centering
  \includegraphics[width=.54\linewidth]{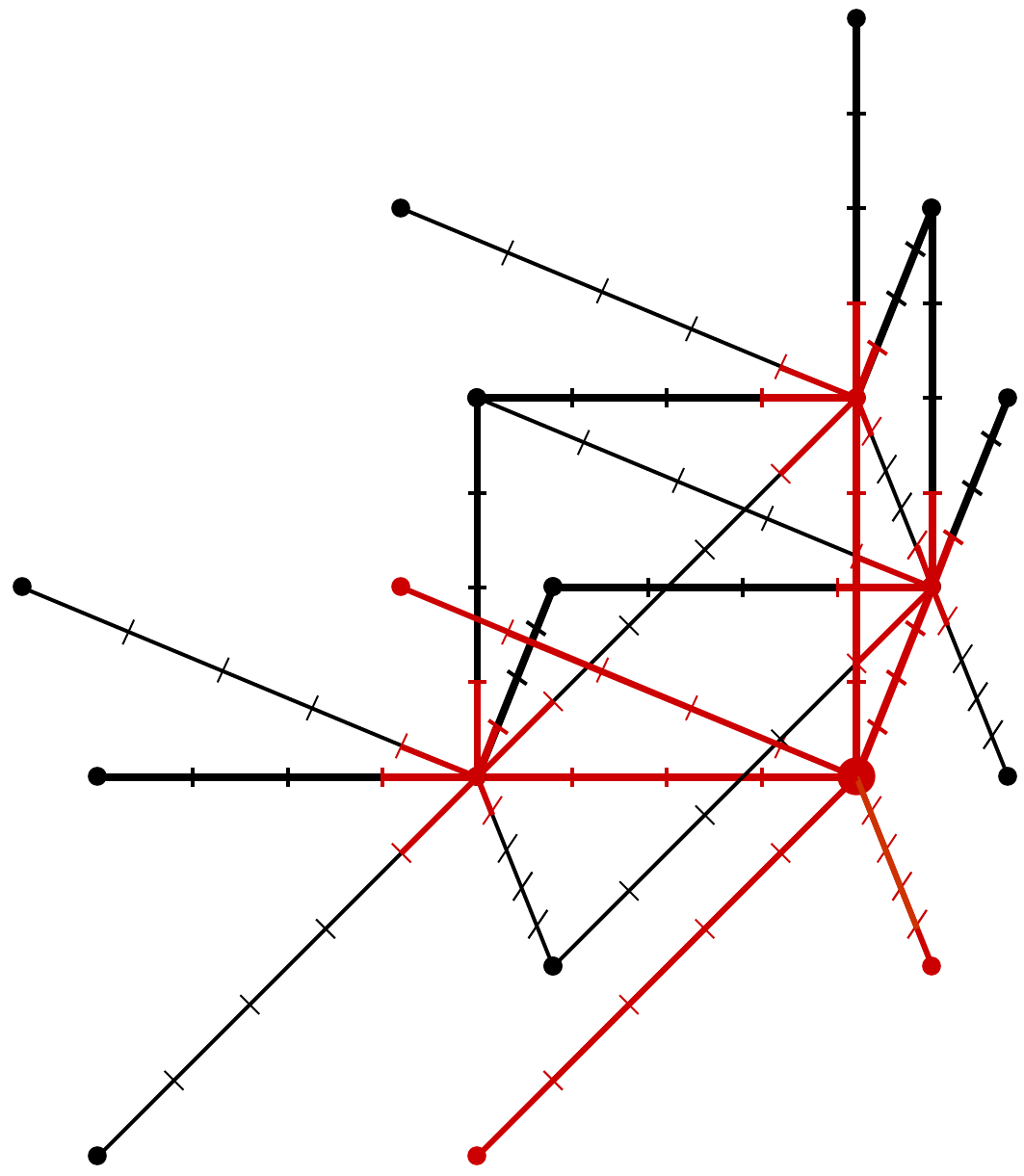}  
  \caption{In the second step, the cluster radius grows by one and it merges with clusters diagonal to the original excitation.}
  \label{fig:step3}
\end{subfigure}
\caption{Two growth steps for weighted UF on a toric code decoder graph with $p=0.8\%$ and weights truncated to the nearest integer (for performance estimates, we truncate to the nearest tenth). Some edges are omitted for clarity. In this case, we have two types of edges: weight four edges (bold) in the cardinal directions, and weight five diagonal edges (thin).  The red highlight indicates the growing cluster.}
\label{fig:weighteduf}
\end{figure}

The UF decoder proceeds in two steps: syndrome validation, which is used to identify a candidate erasure given the syndromes, and peeling, which is used to decode the candidate erasures.  The addition of edge weights changes only the growth step for each cluster during syndrome validation.  In the original algorithm, we would iterate over all boundary vertices of the smallest boundary cluster and grow the incident boundary edges by one-half.  In the weighted algorithm, we first iterate over the boundary edges to identify the smallest boundary edge weight $w_{\text{min}}$, and then again iterate over the boundary edges to grow the radius of the cluster by $w_{\text{min}}$.  Specifically, each edge weight is updated to $w \mapsto w - w_{\text{min}}$. Figure \ref{fig:weighteduf} illustrates this growth step on a weighted graph \footnote{In practice, unerased edges with both endpoints in the cluster are simply removed, as they cannot change the excitation parity of the cluster.}.  We additionally find a minimum-weight spanning tree during peeling, which remains $O(n\alpha(n))$ time when presorting the edges by weight.  However, this only discernibly improves the unweighted UF implementation \cite{li20192d}.

Unfortunately, the inclusion of weighted edges has the potential to increase the runtime of the decoder.  In the unweighted UF algorithm, each edge can participate in a growth step at most twice.  Consequently, for a bounded degree decoder graph, the total complexity of growing the clusters is $O(n)$.  More generally, given edges with real weights $\{w_i\}$ that have a common measure $m$, we can be assured that each edge with weight $w$ participates in a growth step at most $w/m$ times.  However, as $\{w_i\}$ will almost surely have no common measure, we are left with a worst-case upper bound of $O(n^2)$: during each growth step, we iterate through a list of boundary edges of size $O(n)$, and in each iteration we remove at least one edge.  Fortunately, this can be remedied by truncating the $w_i$ to some finite precision $\varepsilon$, ensuring a common measure while incurring a negligible loss in accuracy.  The corresponding weighted UF decoder then has time complexity $O(n\alpha(n) + n/\varepsilon)$, and in the parameter regimes we tested, runs nearly as quickly as the original.

\section{Numerical Simulations}

In this work, we use a standard depolarizing error model parametrized by a single error parameter $p$ (used e.g. in \cite{chamberland2020topological,chamberland2020triangular}).  Our circuits consist of four fundamental noisy gate operations.

\begin{enumerate}
    \item With probability $p$, each idling step (identity gate) is followed by a Pauli error drawn uniformly at random from the set $\{X,Y,Z\}$.
    \item With probability $p$, each two-qubit CNOT gate is followed by a Pauli error drawn uniformly at random from the set $\{I,X,Y,Z\}^{\otimes 2}\backslash (I \otimes I)$.
    \item With probability $2p/3$, intended preparation of $\ket{0}$ or $\ket{+}$ wrongly prepares $\ket{1}$ or $\ket{-}$, respectively.
    \item With probability $2p/3$, a measurement outcome in either the $Z$- or $X$-basis is flipped.
\end{enumerate}

\begin{figure}[htb!]
\begin{subfigure}{.48\linewidth}
  \centering
  \includegraphics[width=\linewidth]{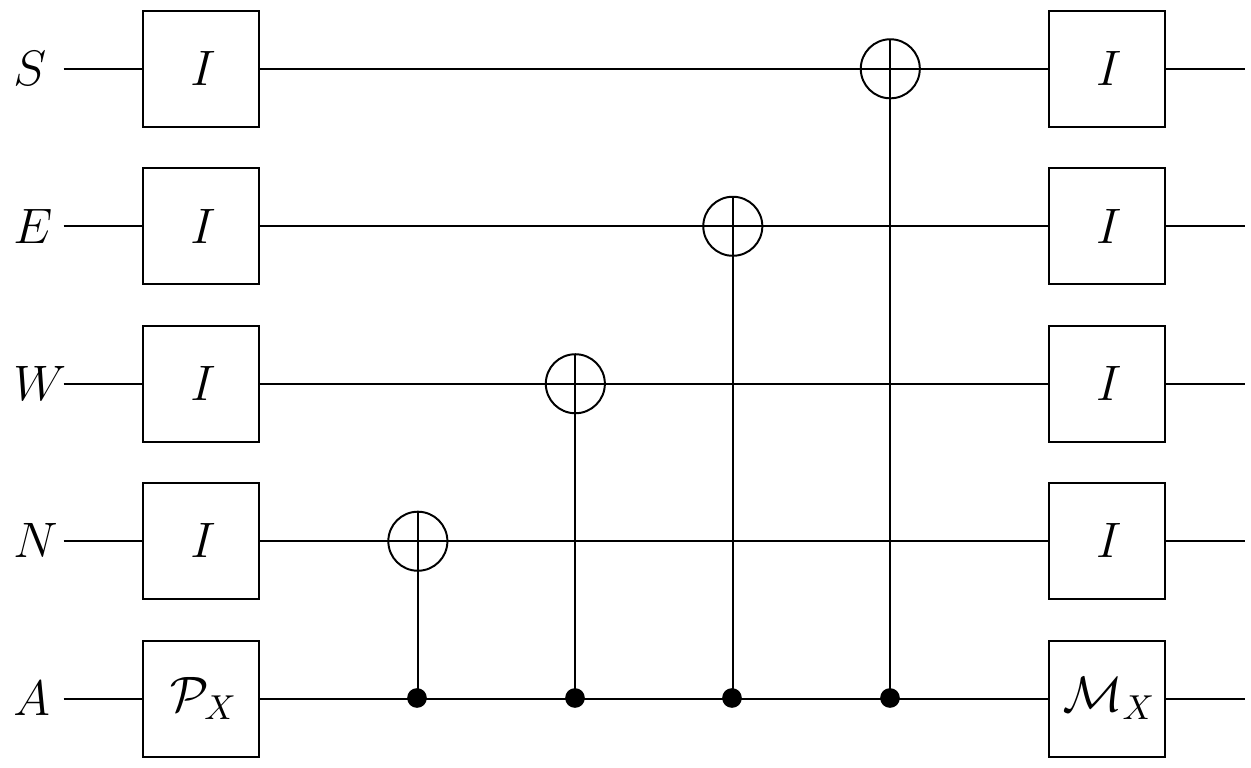}  
  \caption{$X$-stabilizer extraction.}
  \label{fig:Xstab}
\end{subfigure}
\begin{subfigure}{.48\linewidth}
  \centering
  \includegraphics[width=\linewidth]{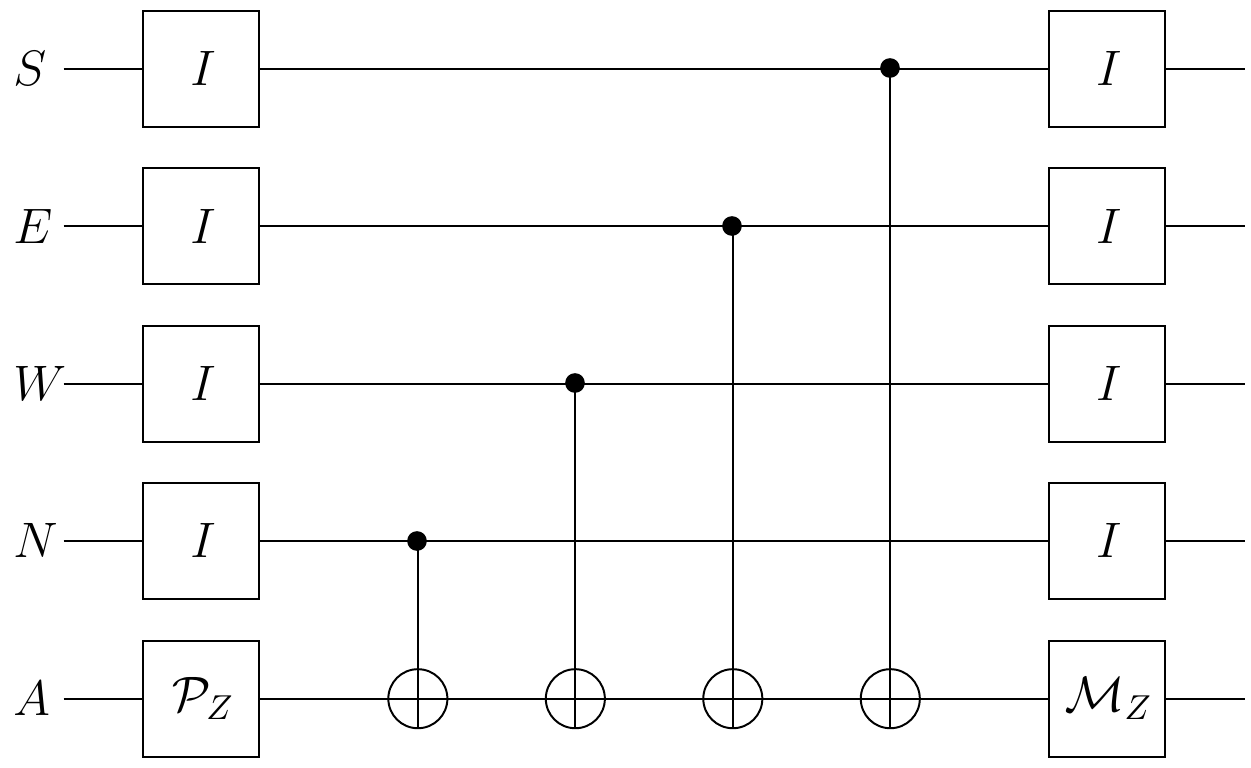}
  \caption{$Z$-stabilizer extraction.}
  \label{fig:Zstab}
\end{subfigure}
\caption{Six-step syndrome extraction on the toric code.  Each ancilla qubit interacts with the data qubit to its north, west, east, and south, in that order.}
\label{fig:extraction}
\end{figure}

Syndrome extraction for the toric code proceeds in six steps: one preparation step, four two-qubit gates, and a measurement step as shown in Figure \ref{fig:extraction}. The decoder graph is formed by connecting all space-time sites that can be jointly excited by a single circuit fault.  Each of these edges is then weighted by $\ln((1-p)/p)$, where $p$ is the sum of the probabilities of those single faults occurring \cite{Dennis:2002,wang2011surface}.

We analyze five different decoding strategies.  We consider MWPM and UF on both weighted and unweighted decoder graphs, as well as UF on a decoder graph with weights truncated to the nearest tenth.  MWPM accuracy has been characterized in a number of works \cite{raussendorf2007fault, Wang:2009, wang2011surface, fowler2012towards, fowler2012timing}; however, as performance depends closely on the microscopic details of the gate and error model, we include it for the sake of direct comparison.  Note that the unweighted decoder graph is not equivalent to a phenomenological decoder graph due to the inclusion of diagonal single circuit-fault edges.  

For MWPM, there have been a number of runtime optimizations \cite{fowler2012towards, fowler2012timing, fowler2015minimum}.  Here, we use a simple localized strategy inspired by \cite{fowler2015minimum} that forms a box around each excitation with dimensions determined by the nearest excitations in the six cardinal directions.  Then, we only check for matchings in which each excitation is matched with another inside its corresponding box.  This simple heuristic speeds up sequential matching, and has performance consistent with previous benchmarks \footnote{Over millions of trials tested at various sizes and error rates, this localized variant always yielded a correct minimum-weight perfect matching.  Matching code provided by IBM.}.  We use Blossom V to perform the matching itself, although we expect additional customization to matching would accelerate the process further \cite{kolmogorov2009blossom}.

We analyze these decoders in three areas: threshold behavior, low error rate behavior, and (serial) efficiency.  The accuracy of the truncated decoder is omitted, as it is indistinguishable from the weighted decoder in the tested regime.  Each trial was decided by performing $d$ rounds of faulty syndrome extraction followed by a terminal perfect round of syndrome extraction.  If any nontrivial logical operator was applied to the two encoded qubits, the trial was declared a failure; else, it was declared a success.

\begin{figure}[htb!]
\begin{subfigure}{\linewidth}
  \centering
  \includegraphics[width=.8\linewidth]{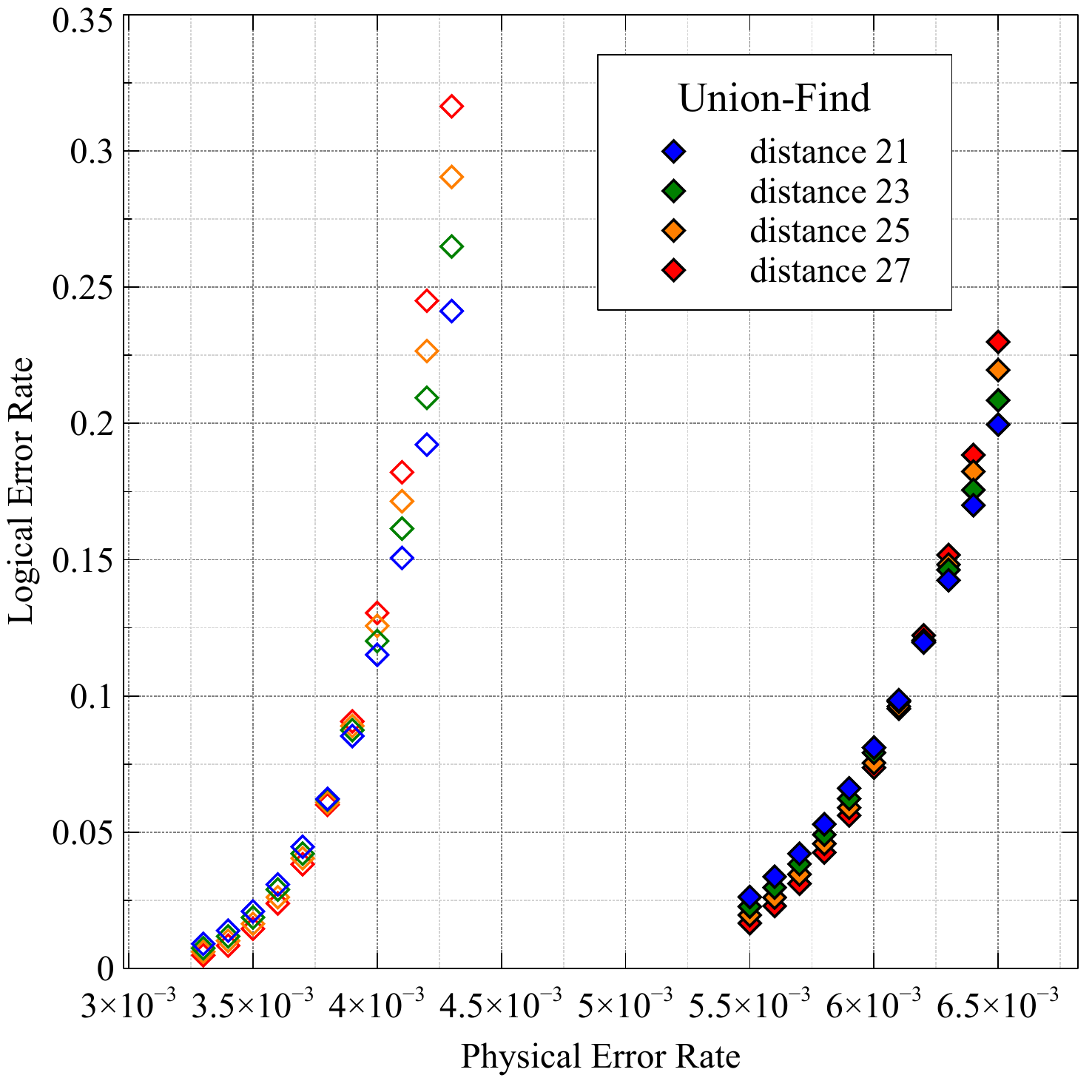}
  \label{fig:uf_threshold}
\end{subfigure}
\begin{subfigure}{\linewidth}
  \centering
  \includegraphics[width=.8\linewidth]{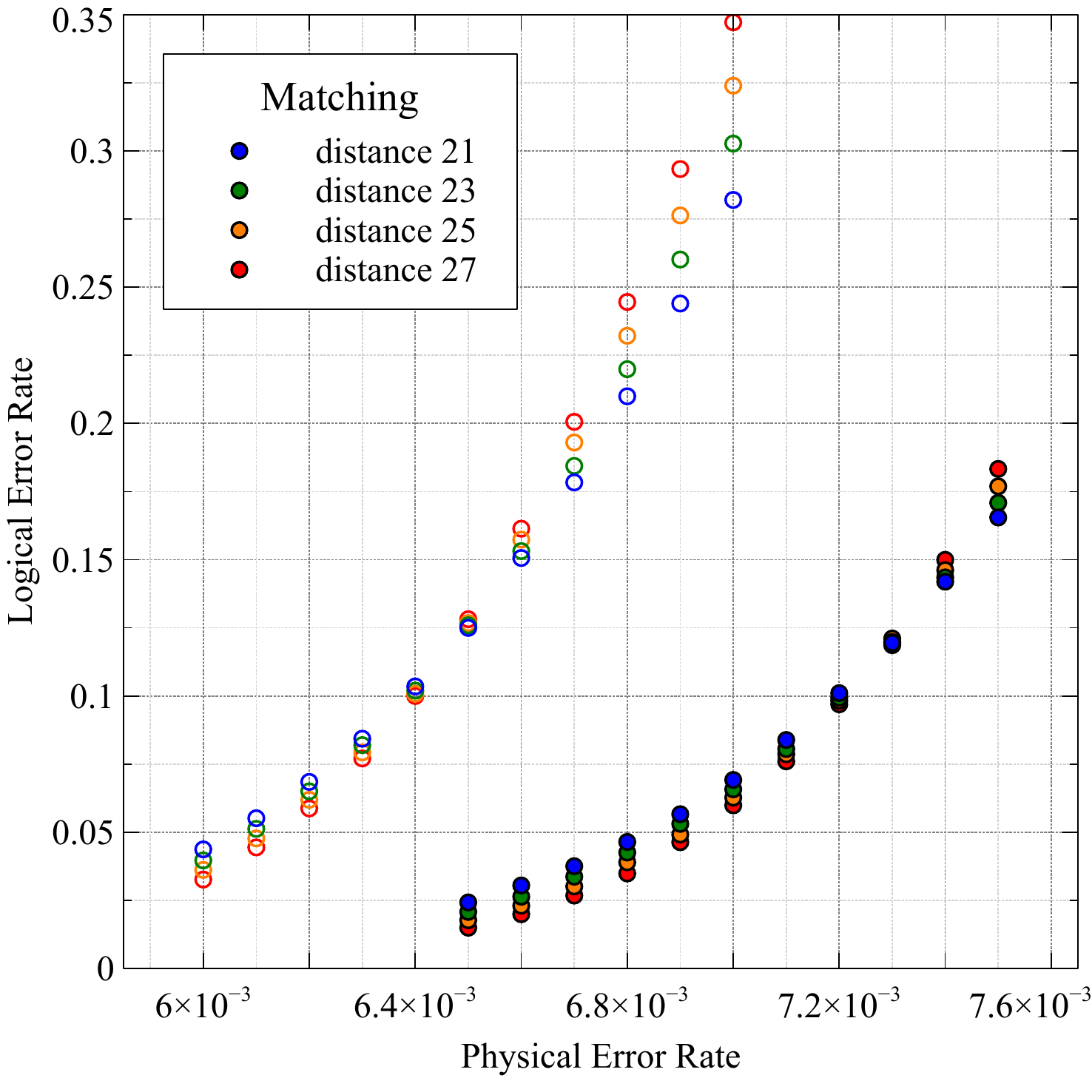}
  \label{fig:m_threshold}
\end{subfigure}
\caption{Threshold behavior for unweighted (empty) and weighted (filled) decoders using both UF (diamonds) and MWPM (circles).  Weighting the UF decoder increases the threshold from $\approx 0.38\%$ to $\approx 0.62\%$, compared with an increase from $\approx 0.65\%$ to $\approx 0.72\%$ for MWPM.  Each point was obtained from $10^6$ trials.  Error bars lie within points, and are omitted throughout.}
\label{fig:threshold}
\end{figure}

Figure \ref{fig:threshold} shows the relative gain in the threshold behavior of weighted versus unweighted decoding.  Weighting the decoder graph significantly improves UF decoding with respect to a less dramatic increase in MWPM.  The threshold for the truncated UF decoder (unshown) approximately matches that of the weighted UF decoder, with a value of $\approx 0.61\%$.  Note that truncation can cause small discontinuities in the logical error rate where the weights jump in value.

We reiterate that these threshold values depend heavily on the specifics of the noise model and operations.  For example, if one assumes quantum non-demolition measurements, the threshold can increase to as high as $\approx 0.90\%$ in the case of MWPM \cite{fowler2012towards}.  In such a model, the weighted UF threshold increases to $\approx 0.76\%$.  However, this model favorably assumes that measurements both report the wrong outcome and project into the wrong eigenstate upon failure, whereas standalone declaration errors can be more damaging.  Thus, it is important to consider the details of the noise model when comparing different absolute threshold estimates.
\begin{figure}[htb!]
\begin{subfigure}{\linewidth}
  \centering
  \includegraphics[width=.8\linewidth]{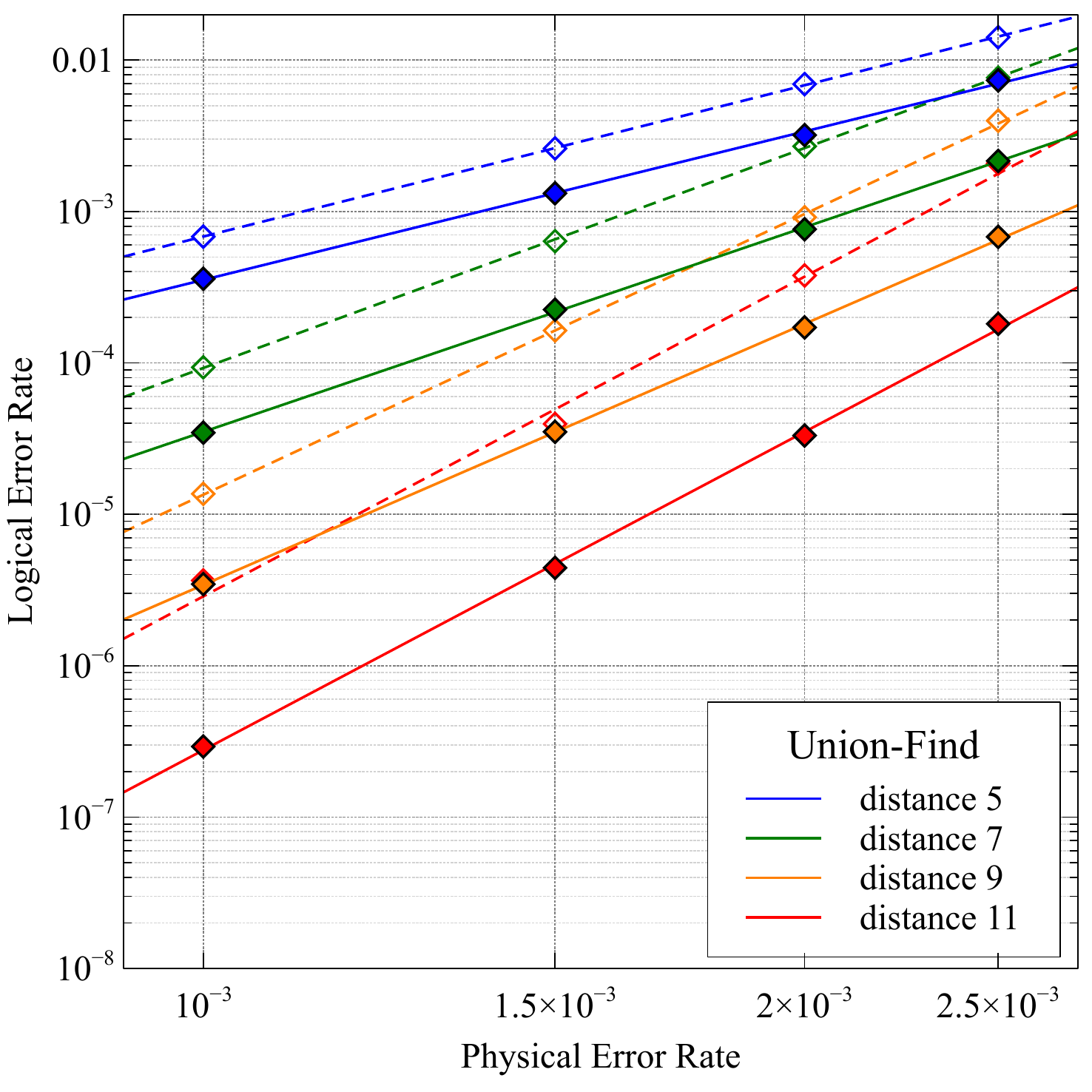}
  \label{fig:uf_low}
\end{subfigure}
\begin{subfigure}{\linewidth}
  \centering
  \includegraphics[width=.8\linewidth]{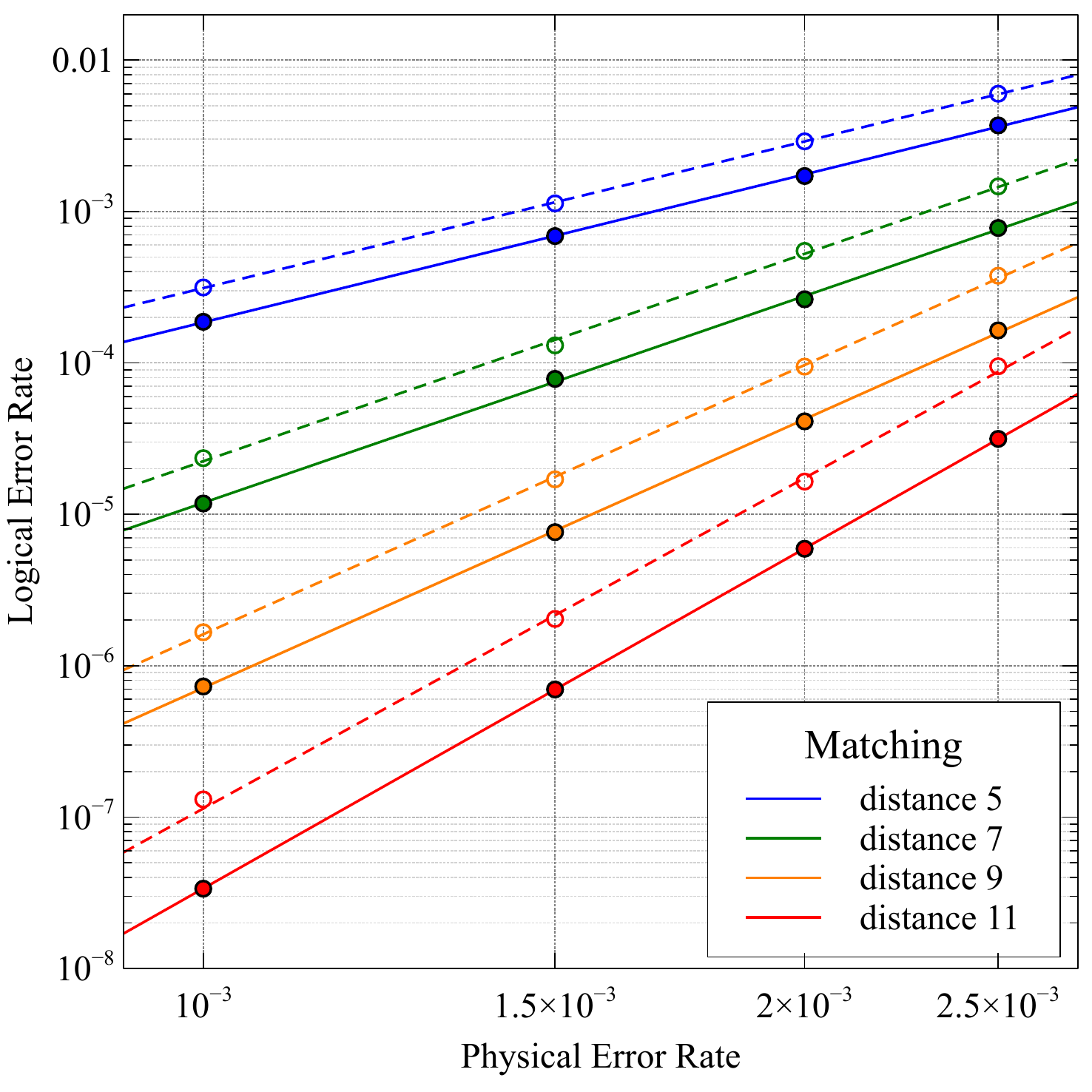}
  \label{fig:m_low}
\end{subfigure}
\caption{Low error rate behavior for unweighted (empty) and weighted (filled) decoders.  Each point was obtained from at least $10^6$ trials and $10^3$ failures.}
\label{fig:low}
\end{figure}

 The low error rate behavior in Figure \ref{fig:low} mirrors that of the threshold behavior.  We observe that weighting UF significantly increases the performance, remaining competitive with matching despite its comparative simplicity and efficiency, even in the fault-tolerance setting.  Table \ref{performance} summarizes the performance comparison between the five considered decoders.

\begin{table}[htb!]
\begin{tabular}{|c|c|c|c|c|c|}
\hline
                                   & $p_{\text{thr}}$ & $\Lambda_{.10\%}$ & $\Lambda_{.15\%}$ & $\Lambda_{.20\%}$ & $\Lambda_{.25\%}$\\ \hline\hline
\textbf{Unweighted UF} & 0.38\%
         & 0.184
            & 0.247           & 0.380            & 0.528        \\ \hline
\textbf{Truncated UF}   & 0.61\%
         & 0.096
            & 0.156          & 0.231            & 0.292                   \\ \hline
\textbf{Weighted UF}   & 0.62\%
         & 0.094
            & 0.151            & 0.219            & 0.292                 \\ \hline
\textbf{Unweighted MWPM}  & 0.65\%
         & 0.075
            & 0.122            & 0.178            & 0.251                      \\ \hline
\textbf{Weighted MWPM}  & 0.72\%
         & 0.057
            & 0.101            & 0.151            & 0.204                     \\ \hline
\end{tabular}
\caption{A summary of the accuracy performance of each decoder.  We approximate the logical performance scaling with $d$ at fixed error rate $p$ as $\propto \Lambda_p^{(d+1)/2}$.  Here, $\Lambda_p$ is estimated by averaging over the three intervals from $d=5$ to $d=11$.}\label{performance}
\end{table}

\begin{figure}[htb!]
\begin{subfigure}{\linewidth}
  \centering
  \includegraphics[width=.8\linewidth]{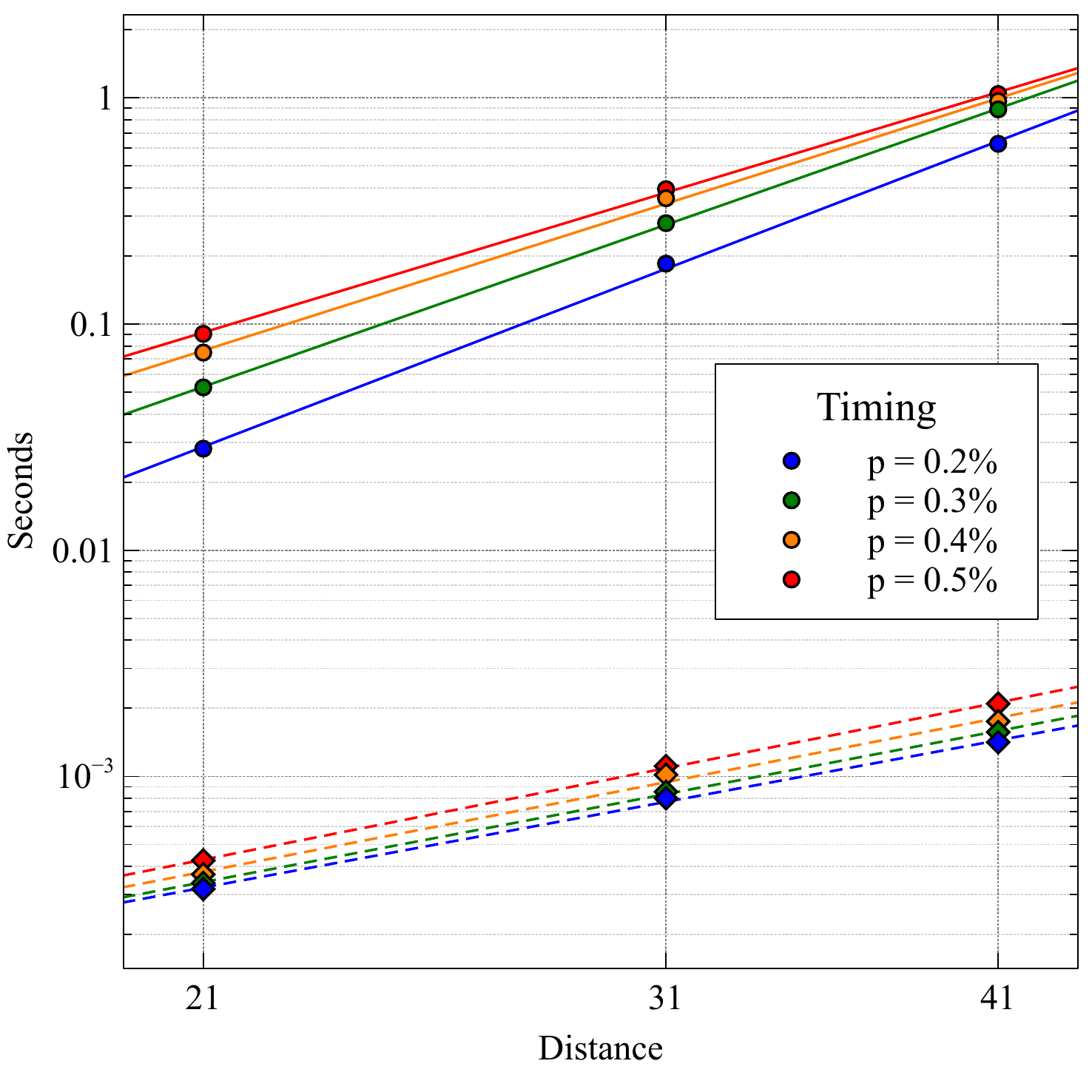}
  \subcaption{Weighted UF (diamonds) scales as $\propto d^{2.2}$, which is nearly linear in $n \propto d^2$. In comparison, our variant of localized matching (circles) empirically runs in time $\propto d^{4.5}$, which is still subquadratic in the total syndrome count.  In total, our implementation of UF runs $\approx10^2-10^3$ times faster than the localized variant of MWPM around the tested regime.}
  \label{fig:normal_timing}
\end{subfigure}
\begin{subfigure}{\linewidth}
  \centering
  \includegraphics[width=.8\linewidth]{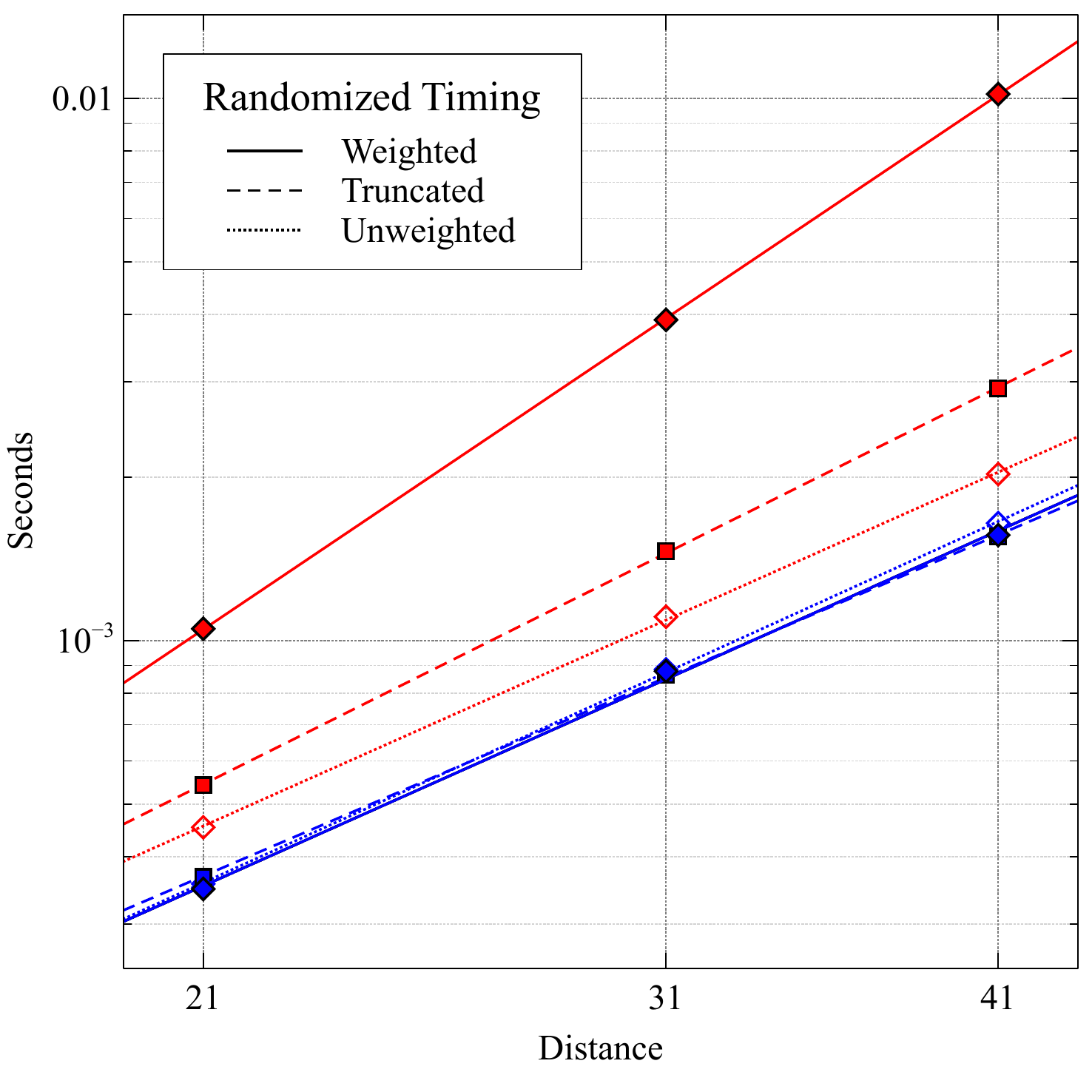}
  \subcaption{Timing on a decoder graph with edge weights $\ln((1-w)/w)$, for $w$ drawn uniformly at random from the range $0.1\%-0.5\%$.  At high physical error rates $p=0.5\%$ (red), weighted UF suffers a significant slowdown, but truncating the weights (squares) recoups most of the unweighted efficiency.  At sufficiently low error rates $p=0.2\%$ (blue), the weighted UF slowdown does not occur.}
  \label{fig:random_timing}
\end{subfigure}
\caption{Timing for weighted MWPM and UF decoders.  Times are reported per extraction cycle, obtained from timing the offline decoders on a $\propto d\times d\times d$ decoding instance and dividing over the $d$ cycles.  For example, a linear time decoder would run in $\propto d^2$ time.  Each point is the average of at least $10^3$ trials on a single $2.9$ GHz Intel Core i9 CPU, and timings are consistent with previous benchmarks \cite{delfosse2017almost}.}
\label{fig:timing}
\end{figure}

Unsurprisingly, sequential UF runs significantly faster than sequential MWPM even when using localizing heuristics.  Figure \ref{fig:timing} shows timing in the case of translation-invariant and non-translation-invariant edge weights.  The slowdown to weighted UF is exacerbated by larger clusters at higher error rates when in the presence of more edge weights.  Practically, one would expect to use a variety of weights tuned according to benchmarks on individual gates, and so this slowdown (from $\propto d^{2.2}$ to $\propto d^{3.4}$) is significant.  Fortunately, simply truncating the edge weights approximately preserves the scaling and performance of weighted UF.  Note also that if the error rate is sufficiently low across the entire lattice, then the slowdown does not occur.  This is likely due to a smaller number of boundary edge weights in any one cluster.

Of course, one should take these offline sequential timings with a grain of salt.  MWPM has enjoyed several refinements that have been empirically shown to reduce the runtime to average linear time at sufficiently low error rates, and in principle to parallelized average $O(1)$ time \cite{fowler2012towards,fowler2012timing, fowler2015minimum}.  In addition, recent work has demonstrated micro-architectures and accelerations that allow for UF decoding in the $\mu$s regime per extraction cycle \cite{das2020scalable, delfosse2020hierarchical}, and extending to a weighted graph could likely be accommodated.  While absolutely comparing runtimes is difficult, we expect that the speed of the UF decoder should ultimately outstrip matching due its local flavor and simplicity.  

\section{Conclusions}

In this work, we benchmarked a weighted variant of the UF decoder in the full fault-tolerance setting, and demonstrated that it performs comparably to matching while preserving the almost-linear runtime of the original.  Although there can be some slowdown, this can be remedied by truncating the edge weights without a significant loss in accuracy.

Compared to the difficult task of building reliable quantum components, one would ideally use decoders that optimize performance, shifting the burden from a quantum problem to a classical one.  However, depending on the size and details of the decoding problem, and given the simplicity, efficiency, and relatively high performance of weighted UF, it might prove a promising avenue towards practical decoding of the surface code.

\section{Acknowledgments}

The authors thank Andrew Cross and Martin Suchara for providing their code for matching, with permission from IBM.  They additionally thank Christopher Chamberland and Nicolas Delfosse for helpful comments.  The computational resources for simulations were provided by the Duke Computing Cluster.  This research was supported in part by the ODNI/IARPA LogiQ program through ARO grant (W911NF-16-1-0082), ARO MURI (W911NF-16-1-0349 and W911NF-18-1-0218), and EPiQC -- an NSF Expedition in Computing (1730104).

%\tableofcontents

\bibliography{bibliography}

\end{document}